\documentstyle[epsf,aps,prl,multicol]{revtex}
\begin{document}
\newcommand{\epsi}{\epsilon}
\newcommand{\eps}{\varepsilon}  
\newcommand{\vare}{\varepsilon}  
\newcommand{\vareps}{\varepsilon}  
\newcommand{\De}{$\Delta$}
\newcommand{\de}{$\delta$}
\newcommand{\mc}{\multicolumn}
\newcommand{\be}{\begin{eqnarray}}
\newcommand{\ee}{\end{eqnarray}}
\newcommand{\einf}{\varepsilon^\infty}
\newcommand{\ez}{\varepsilon^0}  
\newcommand{\nvec}{{\bf \hat{ n}}}  
\newcommand{\Pvec}{{\bf P}}  
\newcommand{\Evec}{{\bf E}}  
\renewcommand{\L}{{\rm L}}
\newcommand{\T}{{\rm T}}
\newcommand{\fine}{\begin{verbatim} -------- \end{verbatim} }
\draft \title{\bf 
Polarization fields in nitride  nanostructures: 
ten points to think about}
\author{Fabio Bernardini and Vincenzo Fiorentini}
\address{Istituto Nazionale per la Fisica della Materia and 
 Dipartimento  di Fisica, Universit\`a di Cagliari, Italy}
\date{\today}
\maketitle

\begin{abstract}
Macroscopic polarization, both of
intrinsic and piezoelectric nature,
is unusually strong in III-V nitrides, and
the built in electric fields in the layers of
 nitride-based nanostructures,  stemming from
polarization changes at heterointerfaces, 
have a major impact on the properties of single and multiple 
quantum wells, high mobility transistors, and thin films.
The concepts involved in the theory and
applications of polarization in nitrides have encountered
some resistance in the field. Here we discuss critically
ten ``propositions'' aimed at clarifying the main 
controversial issues. \\


\end{abstract}


\begin{multicols}{2}

III-V nitrides are a new frontier of semiconductor physics \cite{oa1}
.
One of their basic properties, macroscopic polarization,  offers
unique opportunities for device design and  basic investigations. 
Polarization manifests itself as built-in electrostatic fields in
polarized layers interfaced  to each other.
These fields affect the characteristics,  performance, and
response of multilayer nanostructured devices.
Here  we  discuss some  theoretical basics and a number 
of practical   implications of polarization fields 
for nitride nanostructures.

The microscopic foundations
of dielectric polarization theory,
 established in the first half of the present decade
\cite{Resta,VKS},  identify		
 the zero-field or {\it transverse} polarization $\Pvec_\T$
of a periodic bulk crystal
 as the  gauge-invariant Berry's phase
of its occupied Bloch orbitals,
accumulated in an adiabatic transformation
of  the system from some reference state to its actual state
\cite{Resta,VKS}.
$\Pvec_\T$, which can equivalently be viewed as the integrated
polarization current flowing through the crystal during the
transformation, has no  relation with the charge density of
the polarized dielectric. Importantly, $\Pvec_\T$
can be calculated accurately  from first-principles
density-functional 
calculations \cite{Resta,VKS,noi.pol,noi.prl,noi.off,noi.eps}.
The {\it absolute} polarization of a material can also
be obtained  by  referencing its polarization to
that of a system for which $\Pvec$=0 by symmetry 
 (for  wurtzite, this may
be  the  zincblende phase \cite{noi.pol}). Although 
 not directly measurable, $\Pvec_\T$ is relevant to
polarization and fields in finite, {\it e.g.} multilayer,
 systems  through classical relations of the sort
$\Pvec_\T = \tensor{\eps}\!_0 ~\Pvec_\L$,
where
 $\tensor{\eps}\!_0$ is the (measured or computed
 \cite{noi.prl,noi.eps}) static dielectric tensor,
and $\Pvec_{\rm L}$ is the  dipole moment per unit volume
of the finite system, {\it i.e.}
the termination-dependent
 longitudinal polarization 
(also identified with minus the screened field generated by the net
polarization charge at the sample surfaces or outer interfaces). 
It is simple, as discussed below, to generalize
such expression to a multilayer systems relevant
in technology in terms of the more fundamental 
$\Pvec_{\rm T}$ only \cite{pss}. In the rest of the paper, 
any mention of polarization refers to the transverse variant
$\Pvec_{\rm T}$.

In the case of III-V nitrides \cite{nota},  we deal 
with the total polarization  ${\bf P}={\bf P_{\rm sp}}$ +
 ${\bf P_{\rm pz}}$($\epsilon$) in a given
strain state at zero temperature, in the absence of external
fields. 
A key point is that in wurtzite nitrides the total
polarization at zero strain, known as spontaneous, is non-zero and
large. The spontaneous polarization has a fixed direction
on general grounds \cite{burns}, and specifically it points
along (000$\overline{1}$) in III-V nitrides.
The strain-induced piezoelectric term
${\bf P_{\rm pz}}$($\epsilon$) can in principle 
point in any direction depending on strain.
Since nitride multilayers are  usually grown along the (0001) axis,  
${\bf P_{\rm pz}}$($\epsilon$) actually lies along that axis.
Given  the results of Ref. \cite{noi.pol} and the conventions
explained therein, for a layer under tensile strain 
(lattice constant must expand to fit on substrate, 
{\it e.g.} AlGaN on GaN) 
the piezoelectric polarization points along (000$\overline{1})$,
whereas for a layer under compressive strain 
(lattice constant must contract to fit on 
substrate, say InGaN on GaN) it points along the (0001) direction.
Similarly to ${\bf P_{\rm sp}}$, the piezoelectric  component
${\bf P_{\rm pz}}$($\epsilon$) is also quite unusually large, given
the giant piezoelectric constants of III-V nitrides 
\cite{noi.pol,shimada}.

It is clear that the two components just discussed
 can sum up or cancel each 
 other out depending on strain and polarity; whether this
increases or decreases the field in the layer,  depends on the
polarization and geometry of the neighboring layers.
In fact, the polarization  of a material $A$ manifests itself  
at the interface with a  different medium $B$ as a	
fixed local charge accumulation.   If the interface is  
insulating and gap-states--free (as is generally the case for the
isovalent common-anion nitride interfaces),  simple relations
holds \cite{noi.off,noi.prb}
between interface charges and bulk transverse polarization.
These  were verified directly in ab initio calculations
on interfaces of binary nitrides \cite{noi.prl,noi.off}
and of other materials \cite{noi.eps}:
polarization charge densities of order
10$^{13}$ cm$^{-2}$,    localized in an
interface region $\sim 3-4$ \AA\, thick  are found to 
be common. As a consequence of these large 
charge accumulations  \cite{nota2}, macroscopic
 electrostatic fields  exist  in the interfaced layers,
screened by dielectric response [free-carrier screening
is taken up later on in the paper].

The typical effects of polarization fields are such, as  discussed further 
below, that comparing the  predictions of polarization theory with
typical  experiments is an inherently indirect process,    requiring
considerable modeling. It is essential that this modeling be done
appropriately. We have previously applied  \cite{noi.apl,noi.prb}
a sophisticated (and complicated)  approach
to realistic nanostructures,
whereby we solve selfconsistently
an  empirical tight-binding Schr\"odinger equation 
and Poisson's equation. The latter reads
\begin{equation} 
\frac{d}{dz}D =
\frac{d}{dz}\left(-\varepsilon\frac{d}{dz}V+P_T\right)=
e\left(p-n\right),
\label{eq:1}
\end{equation}
where the position-dependent quantities  $D$,   $\varepsilon$,  $V$,
$p$, $n$, and  $P_T$ are respectively the displacement field,
dielectric constant,  potential, hole density, electron density, and 
 total transverse polarization. The effects of composition,
polarization, and free carrier screening are thus included in full,
and  non-equilibrium as well as equilibrium carrier distributions can
be studied. 

In this paper, instead, we discuss
qualitative aspects of
 polarization effects referring
to a simple  expression for the
 the electrostatic field in
 the  $j$-th layer of
a  periodic but otherwise  arbitrary 
multiple quantum well or  superlattice made of layers of 
materials $k$ of thicknesses $l_k$ and dielectric constants 
$\varepsilon_k$:
\be
\Evec_j =\frac{\sum_k  (\Pvec_k - \Pvec_j) l_k /\eps_k}{\eps_j
\sum_k l_k/\eps_k},
\label{campo}
\ee 
with sums running on all layers (including the $j$-th). $\Pvec_j$ is
the total, 
spontaneous plus piezoelectric, transverse polarization of the material
in layer $j$. 
This expression, which can be derived 
 \cite{pss} under the somewhat restrictive 
assumptions of 
periodic boundary conditions and
 no  free carrier screening,
shows at a glance that polarization effects couple together
all parts of a multilayer structure via polarization differences.
Also, simply using Eq. \ref{campo} and the values calculated in
Ref. \onlinecite{noi.pol} for piezoconstants and intrinsic
polarization, several often-neglected facts are immediately
apparent.\\ 
{\bf 1 --} {\bf Spontaneous and piezoelectric
polarizations  are  
comparable}, and neglecting either of the two components leads
to inaccuracies. This is best seen in 
  numerical examples; for instance,
 Al$_{0.2}$Ga$_{0.8}$N assumed to be strained on GaN  has a piezoelectric
 component of  --7.5 mC/m$^2$ 
and a spontaneous one of    --39.4 mC/m$^2$;
an In$_{0.1}$Ga$_{0.9}$N layer strained on GaN
has  +15.6 mC/m$^2$ piezo, and --29.3 mC/m$^2$ spontaneous.
This  reflects directly on fields. For a
strained  GaN/In$_{0.1}$Ga$_{0.9}$N isolated QW
(given the of GaN spontaneous polarization --29 mC/m$^2$)
 one
gets a field of 30 kV/cm on neglecting the piezo component,
 while the full value is --1.63 MV/cm.
For an  unstrained isolated  Al$_{0.2}$Ga$_{0.8}$N/GaN 
QW, neglect of  spontaneous polarization gives a field of --1.14 MV/cm 
instead of the actual  --1.89 MV/cm.
The effect is less dramatic in the AlGaN/GaN systems since
 the piezoconstants of AlN are extremely large so that
the piezoelectric term is sizable even at low strain.
Of course, besides 
the errors it causes, neglect of either of the components %
is conceptually ill-founded.
\\
{\bf 2 --} {\bf Only polarization differences at interfaces
are relevant}. This fact is a consequence of electrostatics via
$\nabla\cdot\Pvec$=--$\sigma$, it is consistent
with the tenet of basic polarization theory \cite{Resta} that  only  
polarization differences are meaningful, and
is also borne out by Eq. \ref{campo}, which is derived 
under simple electrostatics assumptions.
Some workers incorrectly assumed earlier on,
 that in a generic MQW 
 the field in the active (well) region was
 given by the polarization/dielectric constant
  ratio  of the well material only. In the case of {\it e.g.} 
 a GaN-matched MQW with equally thick layers   of 
GaN and Al$_{0.2}$Ga$_{0.8}$N, this assumption leads to 
predict a field of  --3.2 MV/cm in the well, while
  using the correct procedure the fields are, as expected,
equal and opposite in  the well and barrier, and 0.74 MV/cm in
modulus. \\
{\bf 3 --} {\bf The geometry of the structure determines directly
the fields}. This comes about for two reasons.
First, the field is determined by the structure geometry via
Eq. \ref{campo};
second, the field-induced potential drop changes
with layer thickness.
In a GaN-matched Al$_{0.2}$Ga$_{0.8}$N/GaN 
system with 30 \AA\, wells and 60 \AA\, barriers the
field in the wells is --1.31 MV/cm with a potential drop
of 0.39 eV; 
if the wells are  enlarged to 60 \AA, the well field is
--0.98 MV/cm, and the  potential drop is --0.59 eV.
Hence, doubling the thickness of a single 
layer does not straightforwardly produce a doubled
potential drop. This is most relevant since it is
just  this potential drop that causes 
the optical shifts  visible in PL measurements.
Note in passing that the number of barrier/well units 
and their growth sequence is also  relevant.
\\
{\bf 4 --} {\bf Piezo vs spontaneous dominance}
depends on composition.
 Nanostructures involving AlGaN/GaN will be dominated by
spontaneous fields,   due to the huge difference in spontaneous
polarization between AlN and GaN, and their relatively small lattice
mismatch.  GaN/InGaN nanostructures will be dominated by
piezoelectricity, since the spontaneous polarization  
difference between InN and GaN  is small, whereas
their lattice mismatch is large.
 This becomes obvious considering that, in the
Vegard hypotesis,   for each 10 \% molar fraction change
the specific strain
 is --1 \% for InN in GaN, and +0.25 \% for AlN in GaN, while
  $\Delta \Pvec_{\rm sp}$ is  +0.3 mC/m$^2$ and --5.2 mC/m$^2$
for the same cases.\\ 

\begin{figure}
\epsfclipon
\epsfxsize=7.5cm
\narrowtext
\centerline{\epsffile{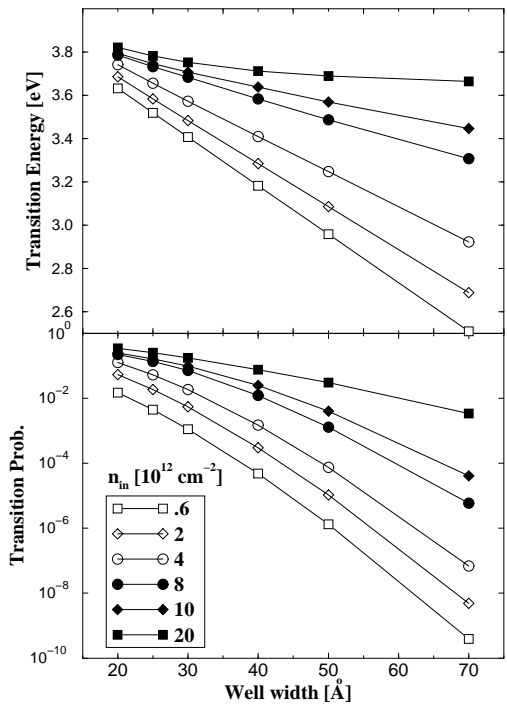}}
\caption{ Transition energies and probabilities of an
isolated 
Al$_{0.2}$Ga$_{0.8}$N/GaN quantum well as function of well thickness 
and excitation density (see Ref.\protect\cite{noi.prb})}
\label{fig.prb}
\end{figure}

The interplay of free carriers (intrinsic,
extrinsic, injected, photogenerated, etc.)
and polarization fields in nitride multilayers
turns out to give rise to a rich variety of new phenomena.
A closer look to these issues
 through accurate simulations
and thoughtful experiments has indeed revealed several 
hot spots:\\
{\bf 5 --}  {\bf Optical transitions are red-shifted, slowed down, 
and made less intense by the fields}.
This effect, due to the progressive spatial
separation of hole and electrons at the two sides
of the QW, 
 was observed by several workers in strained
GaN/InGaN \cite{exp-red.in},
and strained as well as unstrained AlGaN/GaN MQWs 
\cite{exp-red.al,langer}. In calculations, it was shown 
that at excitation densities typical
for PL measurements, the shift in a single 
QW is perfectly linear with the thickness, and
the oscillator strength is suppressed accordingly
\cite{noi.apl,noi.prb,peng}. The shifts are most often such that 
the transition occurs well {\it below} the band gap energy.
Identifying the magnitude of the polarization via measurement of
the shifts requires an accurate modeling, expecially 
when different well thicknesses are used
at the same time in a MQW \cite{langer}, and also in the
small-thickness limit where the field-induced red shift 
is not linear due to the competing  confinement blue shift. \\
{\bf 6 --} {\bf Optical excitation, electrical injection, or doping,
lead to partial screening} of the fields due to generation of
  free carriers in the active layers or thereabouts. This was 
observed in experiment as a blue shift 
 of the transition energy and recovering oscillator strength as a
function of excitation power \cite{noi.apl,exp-red.in}. 
As indicated theoretically \cite{noi.apl,noi.prb},
 the blue shift is in fact relative to the field--red-shifted energy, 
and  failure to recognize the red shift as due 
to polarization  has lead to overestimates of the In contents in
InGaN layers. Referring
to Fig. \ref{fig.prb}, more points are worth a mention. 
A very substantial 
excitation is needed to recover appreciable oscillator strengths, 
and to blue-shift the transition back towards the flat-band value; 
this finding agrees with thresholds of GaN-based lasers being up to 
one order of magnitude larger than in GaAs-based devices. 
Further, the oscillator strength never
reaches unity, and the energy never equals the gap plus confinement
energy  even in the highest power limit. This
 is because  the relevant  length scale  for free carrier
 screening of the field is of order 15-20 \AA\, due 
 to the larger spatial extension of the quantum-confined carriers
as compared to the polarization charge. A possible
 exception are  extremely
thin wells where the screening length is similar to the well width,
and  the confinement blue shift competes with the field red shift.  
Note  that while
 excitation-driven screening dies out as excitation is removed,
permanent screening can be and has been achieved by remote 
doping \cite{noi.prb,doping}.\\
 {\bf 7 --}  {\bf  The field-induced potential drop can become
similar to the gap} of the layer material, for  typical fields
(several MV/cm) over typical well thickness ($\sim$ 100 \AA). 
Screening by the intrinsic  
carriers becomes important in this limit. As could be guessed,
simulations show that at non-zero temperature,
as a result of free carrier generation at the opposite 
ends of the sample, the field-induced potential drop across a given layer 
never  exceeds the gap, but rather \cite{noi.prb} 
  the field obeys $|\Evec| = E_{\rm gap}/d $ for 
thicknesses  $d \geq d_c =E_{\rm gap}/|\Evec|$. 
Since the  relevant $d_c$'s are
generally larger than typical QW thicknesses,  the gap
closure regime is generally not of interest there, but it may be
to thick layers (used {\it e.g.} to measure alloy properties).
On the other hand, cases exist where intrinsic carriers
determine the boundary conditions, as in the next point.\\
 {\bf 8 --} {\bf Self-doping} may occur 
in ``constrained'' systems. This was observed experimentally
by at least three groups \cite{bik,bik2,oa} 
and confirmed theoretically \cite{za} in the instance of
 high mobility transistors. We pick the example of a thick
($\sim$ 300 \AA)  AlGaN layer on a very thick 
($\sim$ 1.5 $\mu$m) GaN buffer,  contacted at the top. The metal
Fermi level, the  Schottky contact barriers, and  the interface band
offset limit the maximum admissible potential drop through the AlGaN
layer    to about 1 eV. The polarization  field would instead 
imply  a drop of about 4
eV. For equilibrium to be established, a large part of the polarization
charge at the heterointerface  must therefore 
be screened by free carriers, coming in
from the GaN layer (not from the  contact due to 
the Schottky barrier) following the strongly attractive polarization-induced
potential drop  on the GaN side. The GaN layer is typically
unintentionally $n-$doped to at least 10$^{16}$ cm$^{-3}$, hence a
fair source of electrons. The end result at equilibrium  \cite{bik,bik2,oa,za}
is the accumulation of a
extremely high-density (up to 10$^{13}$ cm$^{-2}$) 
two-dimensional electron gas at  the AlGaN/GaN
heterointerface channel  (the mechanism is
 summarized, {\it e.g.}, in Ref. \cite{noi.prb} and
discussed in detail in  Refs. \cite{bik}, \cite{oa}, and \cite{za}).
The mobility and channel transconductance 
are increased accordingly \cite{za}, also thanks
 to the absence of ionized impurities in the channel itself.
 Two groups\cite{bik,bik2} have interpreted their results in
terms of piezoelectricity only, while a third \cite{oa}
has included spontaneous
polarization. At the level of charge density CV profiling
measurements, the uncertainties are sufficiently large, and the 
interpretations indirect enough that both ways of proceeding
can get acceptable credit; as mentioned in the next point,
however, only  spontaneous polarization  stands the 
consistency check with the sample  polarity.

\begin{figure}
\epsfclipon
\epsfxsize=7.5cm
\narrowtext
\centerline{\epsffile{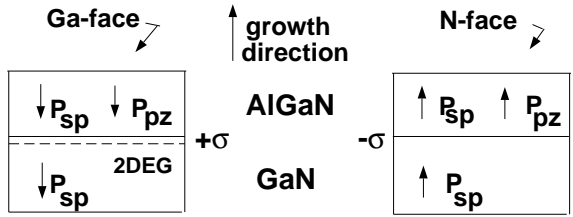}}
\caption{Interplay of polarity and 2DEG formation in the
 HEMT channel (adapted from Ref. \protect\cite{oa}).}
\label{fig.oa}
\end{figure}

Two-dimensional {\it hole} gases
can also be obtained thanks to fields, 
as demonstrated recently \cite{db}.
A superlattice is  designed so that the field in the doped
barriers and well is equal and opposite; this is
achieved to a good approximation with a MQW with equal well
 and barrier thickness ( see Eq. \ref{campo}). The field enhances,
essentially by field ionization, the  release of holes
  from the relatively deep acceptors in the GaN or alloy
layer, and funnels the holes to the interface holding negative
polarization charge. Holes accumulate at the interface
 forming a degenerate 2-D gas. Spontaneous polarization
was accounted for, and measured charge densities matched well those
predicted by modeling.\\ 
 {\bf 9 --} {\bf Polarity plays a key role}:
since the polarization in each layer has  a definite sign
dictated by strain and by the spontaneous polarization direction, the
combination of polarizations to produce fields (Eq. \ref{campo}) is
 determined by the   polarity of the  structure. This is in turn
determined by the substrate-  and  preparation-dependent growth face
[N-face, {\it i.e.} (000$\overline{1}$) surface up,  or rather Ga-face,
{\it i.e.} (0001) surface up].  Hence control on, and awareness of the
polarity issue is also essential, since 
it is easy to envisage how polarity influences experiments. 
The  measurement of polarization effects
through  field-induced optical shifts  
in quantum wells is ``sign-blind'', since the field sign is 
immaterial to the observed 
shift. In electrical devices such as HEMTs, if a charge
accumulation is expected at a given interface for a given
 polarity, the occurrence of  the   opposite polarity will cause
instead a depletion, as displayed in Fig. \ref{fig.oa}. 
Of course, the magnitude of polarizations in the layers 
will determine the density of the 2-D gas, hence quantitative
agreement  can be expected only including both components.
Indeed, in CV experiments on controlled-polarity
HEMT structures \cite{oa}, the appearance (or non-appearance)
 of a 2-D gas at a specific  interface was shown to correlate
\cite{oa}  with the existence of the spontaneous  polarization
component, and its value was estimated in fair agreement with our
calculations \cite{noi.pol}.\\
{\bf 10 --} {\bf Piezoelectric and spontaneous polarization can be
independently controlled} to a large extent, basically because only 
the first depends on strain. It is easy to show  \cite{noi.prb}
that it is possible to obtain completely unstrained, yet polarized 
layers, as well as  strained yet unpolarized systems. In the first
case, the piezo component vanishes by construction, hence spontaneous
polarization is unambiguously the  only remaining source of field, and
can therefore be measured. In the second case, the composition must be
such that polarization differences  are zero across all interfaces,
whence  the fields vanish in all layers. It is rather clear that,
since realizing this situation requires compensating piezo and
spontaneous components,   alloying with Al {\it and} In will be
needed. For  instance, the field-free case should be obtained with
GaN-matched GaN/Al$_{0.10}$In$_{0.06}$Ga$_{0.84}$N  MQWs with equally
thick wells and barriers. Also GaN-matched
GaN/Al$_{0.68}$In$_{0.32}$N  MQWs should exhibit the same behavior. 
In practice, one would measure the red shift vs. thickness
for a series of GaN/AlInGaN MQWs starting at null In content;
upon   progressively increasing the In content, the shift should be
first seen to vanish, and then to reappear for larger In concentration.
Problems (or opportunities) 
that should be watched out for in modeling are that
the interface band offset   may switch to type-II, and that alloy
phase separation is likely.
Of course, the ``critical'' concentrations mentioned above
 must be taken as indicative only, in view of their sensitivity
to  the intricate interdependencies of strain,
lattice parameters, piezoconstants, and spontaneous polarization
 values.

We thank Oliver Ambacher for
discussions and for pointing out recent experimental work
in the field.
VF thanks the Alexander von Humboldt Foundation for support of his
stays at the WSI. 
Support by INFM under a Section-E PAISS project is acknowledged.

\end{multicols}



\begin{thebibliography}{99}
\bibitem{nota2}
Not related to dangling bonds
as in abrupt heterovalent junctions, and  not implying
interface  instability  \cite{noi.off}.
\end{thebibliography}
\end{document}